# Terahertz-based attosecond metrology of relativistic electron beams

R. K. Li[1*†], M. C. Hoffmann[1*], E. A. Nanni[1*], S. H. Glenzer[1,2], A. M. Lindenberg[2,3,4], B. K. Ofori-Okai[1,2], A. H. Reid[1], X. Shen[1], S. P. Weathersby[1], J. Yang[1], M. Zajac[4] and X. J. Wang[1†]

[1]SLAC National Accelerator Laboratory, 2575 Sand Hill Road, Menlo Park, California, 94025, USA

[2]PULSE Institute, SLAC National Accelerator Laboratory, Menlo Park, CA 94025, USA

[3]Stanford Institute for Materials and Energy Sciences, SLAC National Accelerator Laboratory, Menlo Park, CA 94025, USA

[4]Department of Materials Science and Engineering, Stanford University, Stanford, CA 94305, USA

[*] These authors contribute equally.

[†] Corresponding authors: Renkai Li (lrk@slac.stanford.edu) and Xijie Wang (wangxj@slac.stanford.edu)

**Photons, electrons, and their interplay are at the heart of photonic devices and modern instruments for ultrafast science [1-10]. Nowadays, electron beams of the highest intensity and brightness are created by photoemission with short laser pulses, and then accelerated and manipulated using GHz radiofrequency electromagnetic fields. The electron beams are utilized to directly map photoinduced dynamics with ultrafast electron scattering techniques, or further engaged for coherent radiation production at up to hard X-ray wavelengths [11-13]. The push towards improved timing precision between the electron beams and pump optical pulses though, has been stalled at the few tens of femtosecond level, due to technical challenges with synchronizing the high power rf fields with optical sources. Here, we demonstrate attosecond electron metrology using laser-generated single-cycle THz radiation, which is intrinsically phase locked to the optical drive pulses, to manipulate multi-MeV relativistic electron beams. Control and single-shot characterization of bright electron beams at this unprecedented level open up many new opportunities for atomic visualization.**

Challenges in capturing the fastest atomic motion with electron probe beams consist of not only creating femtosecond and shorter pulse durations, but also controlling or characterizing the timing of the probe electron beam relative to pump optical pulses with the same or better precision. High power GHz radiofrequency sources and structures developed and perfected in past decades have been the workhorse for manipulating electron beams, ranging from compression of keV electron bunches for table-top ultrafast electron diffraction and imaging [14-17] to acceleration and temporal characterization of GeV electron beams in kilometer-long X-ray free electron lasers (XFELs). An inherent limitation of the rf approach, however, is associated with the technical challenge to further improve the timing synchronization between high power rf sources and pump lasers. Although rf technology allows generation and characterization of few-femtosecond electron beams [18-19], a viable path to reduce their timing relative to pump optical pulses to below a few tens of femtosecond does not yet exist.

A most promising way to address the timing challenge for pump probe experiments at a fundamental level, is to manipulate the electron beam using electromagnetic radiation intrinsically synchronized with pump lasers. THz radiation generated by optical rectification or other nonlinear processes is locked in time relative to the driving laser pulses with sub-fs accuracy, and thus is an ideal choice. Meanwhile, THz radiation, having ~$10^2$-$10^3$ times higher frequency than rf fields, is much more efficient in shaping the electron beam's temporal characteristics. The manipulations usually involve the introduction of an energy-to-time correlation in beam phase space for pulse compression, or a divergence (transverse momentum)-to-time correlation for streaking and bunch length characterization. To obtain the same amount of correlation in beam phase space, the required change of energy or momentum scales inversely proportionally with the radiation frequency.

Laser-generated THz radiation has recently emerged as a new tool for controlling electrons with high temporal precision. For instance, it has been employed for active switching and spectroscopic mapping of photo-emitted electrons from solids, providing rich insight into carrier behavior in strong near-fields [20]. From isolated gas atoms, photoelectrons ionized by XUV to hard X-ray pulses can be streaked by laser-generated THz radiation into an energy spectrum, which allows characterization of the temporal structure and timing jitter of those ionizing pulses with femtosecond resolution [21-22]. Besides extracting eV-level-energy photoelectrons from nanotips and atoms through field emission, THz acceleration [23-25] and manipulation [26-27] of in-vacuum free electrons with significantly higher kinetic energies will open up a new era for beam physics and ultrafast science. In a few recent demonstrations, sub-relativistic, <100 keV kinetic energy electron beams from DC sources were compressed using laser-generated THz fields to tens of femtoseconds, with their timing jitter stabilized to a few femtoseconds as characterized by THz streaking [28-29].

There is strong incentive to extend THz control to electron beams of significantly higher bunch charge and multi-MeV relativistic kinetic energy and towards the attosecond regime. These beams, featuring extremely high beam brightness, are generated by state-of-the-art high gradient rf sources. High kinetic energy is very effective in suppressing space charge effects for creating and preserving high beam brightness, enabling single-shot measurement of irreversible dynamics in ultrafast diffraction and imaging, as well as studying systems of very low density in the gas phase [30]. However, THz control of bright relativistic electron beams with femtosecond or better precision has not yet been demonstrated, due to the technical challenges of generating and injecting multi-MeV electron beams from an rf gun with pulse durations and timing jitter both significantly smaller than the THz wavelength.

In this Letter, we report on the experimental demonstration of manipulation of bright relativistic electron beams using laser-generated THz radiation, where the timing between the electron beams and the optical pulses can be determined with attosecond accuracy for the first time. The layout of the experiment is illustrated in Fig. 1. A quasi-single-cycle THz pulse (see details in the Methods Section) with up to 10 uJ pulse energy is generated by optical rectification from a $LiNbO_3$ crystal [31-33]. The THz radiation is linearly polarized along the vertical direction and focused by a 3-inch-focal-length, 90° off-axis parabolic mirror. A relativistic electron beam with 3.1 MeV kinetic energy, 20 fs rms pulse duration, and 3 fC ($2 \times 10^4$ electrons) bunch charge is generated from a photocathode rf gun, injected through an aperture in the parabolic mirror and propagates collinearly with the focused THz pulse. The driving laser pulses for the electron-beam and THz generation are split from a common laser pulse, so that the time delay between the THz and electron beams can be controlled by adjusting the optical path length, leaving the rf-induced timing jitter between THz and electron pulses as the only synchronization uncertainty. The THz and electron beams are adjusted so that they have spatial and temporal overlap at the THz focus where the field strength is maximum and the Lorentz force is strongest. The field of the THz pulse at the focus is measured independently in the time-domain by electro-optical (EO) sampling [34].

The electron beam experiences the Lorentz force from the co-propagating THz radiation as $\boldsymbol{F} = -e(\boldsymbol{E} + c\boldsymbol{\beta} \times \boldsymbol{B})$, where $\boldsymbol{E}$ and $\boldsymbol{B}$ are the electric and magnetic components of the THz pulse, $\boldsymbol{\beta}$ is the normalized velocity of electrons, and $e$ and $c$ are the elementary charge and speed of light in vacuum, respectively. The relativistic electron beams with 3.1 MeV kinetic energy travel at 99.0% of the speed of light, and thus the electric- and magnetic-force components almost totally counter-balance each other with only 1.0% residual. This fact further illuminates the challenges of THz control of relativistic electrons compared to non-relativistic ones for which the net Lorentz force is much stronger.

The Lorentz force integrated over the entire interaction region results in a time-dependent transverse momentum kick to the electrons. The time dependence is determined by the temporal profile of the THz radiation and the slippage between the THz and electron beams. The THz-induced momentum-to-time $p_y - t$ correlation within the electron beams translates into a spatial displacement over a free-drift distance, which is measured by a two-dimensional imaging detector located 3.2 meters downstream of the THz slit structure. In Fig. 2e, we show the measured centroid position, with exemplary unprocessed data shown in Fig. 2a-d, of the electron beam for different optical path length delay between the electron beam and THz driving laser pulses. Referring to this correlation calibration curve, one can then map each electron pulse on a shot-by-shot basis onto the time axis using its centroid position. The accuracy of the timing determination is defined as $\sigma_{y0}/D$, where $\sigma_{y0} = 3.0$ µrad root-mean-square is the pointing stability of the electron beam when the THz is not present, and the correlation coefficient $D$ is the slope of the calibration curve. With $D = 7.4$ μrad/fs, the timing accuracy is 0.41 femtosecond rms when we inject the electron beam around $t = 0$ ps in Fig. 2e where the THz-induced streaking is strongest, and at the same time the quasi-linear time window (-0.30 to 0.30 ps) is significantly wider than the timing jitter of the electron beams.

The motion of the relativistic electrons driven by the THz radiation and the induced fields in the slit structure is elucidated by the simulation of the electromagnetic-field evolution in space and time (see Methods Sections). The measured and simulated angular positions of the relativistic electrons at different time delays are shown in Fig. 3a. The results match remarkably well between -0.3 to 0.3 ps when the streaking slope is largest. The simulation results show faster changes at other time delays with the discrepancy due to the fact that the EO measurement of the THz field is inaccurate for very low frequencies, thus limiting the accuracy of our particle motion simulation.

The electric and magnetic components of the Lorentz force experienced by electrons at two different time delays which end up with the maximum and zero integrated angular kicks are shown in Fig. 3 c and e, respectively. The 100-um-thick slit structure is indicated by the two dashed lines. As the THz radiation and the electrons co-propagate in vacuum and far away from the slit, the net Lorentz force is weak because the strong $E$ and $B$ components cancel each other almost completely. Within a distance of a few THz wavelengths from the slit, the reflection results in a different enhancement of the $E$ and $B$ fields and introduces a significant phase shift between the two components, which leads to strong net Lorentz forces. The slit effectively behaves as an open terminal on a transmission line allowing for an enhancement of the electric field while requiring a minimum in the magnetic field. The simulations confirm this interpretation showing that although the particles do see a net kick from both E and B fields, it is the enhanced electric field on the back face of the slit ($z$=0.1 mm) that results in the net integrated deflection, as shown in Fig. 3b.

The centroid positions of the electron beams provide relative-to-laser timing information with high precision, and the profile of the streaked beams distribution allows single-shot determination of the absolute bunch length with femtosecond resolution. In Fig. 4a and b, we show two single-shot snapshots of electron beam profiles with the THz turned off and on, respectively. With the position-to-time correlation curve shown in Fig. 2a and deconvolution with the non-streaked beam profile, it is straightforward to convert the streaked beam profile shown in Fig. 4b into its distribution in the time-domain. The resolution of the measurement, related to the finite size of the un-streaked beams, is evaluated to be 2.5 fs rms (see Methods Section for detail). Further improvement of the resolution can be achieved by reducing the beam size, increasing the THz strength, and utilizing interaction structures with larger field enhancement.

The demonstrated THz metrology with attosecond accuracy constitutes a powerful new tool for direct characterization and enables significant improvement of the overall temporal resolution of ultrafast electron scattering measurements. In Fig. 4 d and f, we show the measured relative-to-laser timing and absolute bunch length of 1000 consecutive shots of electron beams. The shot-to-shot timing jitter is 45.8 fs rms and the bunch length is $21.3 \pm 1.3$ fs rms. The measured timing jitter of each and every electron probe bunch, i.e. the time stamp for each single-shot electron scattering images, can be used to resort the images with respect to the pump laser with sub-femtosecond accuracy, and thus significantly improve the overall temporal resolution, similar to the x-ray-optical cross-correlation technique developed at XFELs [35-36]. The absolute bunch length can be further improved by at least one order of magnitude with the addition of an rf [19] or THz compressor.

In summary, we report on an experimental demonstration of streaking relativistic bright electron beams using a laser-generated THz pulse, which allows determination of the electron beam-to-laser timing with unprecedented sub-femtosecond accuracy, and at the same time direct measurement of the bunch length with a resolution approaching single femtosecond. The result is a major advance in manipulating high energy electrons using THz radiation, which is significantly faster than the traditional rf technology and holds tremendous potential in creating and characterizing isolated electron beams into the attosecond regime [37-40]. Moreover, a distinct advantage of utilizing laser-generated THz is that the manipulation of electron beams is intrinsically synchronized to the driving laser, which essentially eliminates the timing-jitter challenge in pump-probe ultrafast electron scattering measurements and external injection in laser-driven accelerators [41-43]. Utilizing such dramatically improved temporal resolution, one can explore the intriguing opportunity of bright electron-beam based spatiotemporal mapping of the THz electromagnetic fields in metamaterial devices [44] and optically excited wakefields in plasma and dielectric structures with nanometer and sub-femtosecond resolutions. With stronger THz radiation [32] and more efficient interaction structures [25], THz metrology can be extended to GeV-kinetic-energy electron beams in FEL drivers.

**Acknowledgement**

The authors gratefully acknowledge the technical support from SLAC Accelerator Directorate, Technology Innovation Directorate and LCLS Laser Science and Technology Division. This work was supported by the U.S. Department of Energy (DOE) under Contract No. DE-AC02-76SF00515, the DOE Fusion Energy Sciences under FWP 100182, the DOE Basic Energy Sciences (BES) Accelerator and Detector R&D program and the SLAC UED/UEM Initiative Program Development Fund. A.M.L. acknowledges support by the DOE, Office of Science, BES, Materials Sciences and Engineering Division. M.C.H. is supported by the DOE, Office of Science, BES, under Award No. 2015-SLAC-100238-Funding.

**Author contributions**

R.K.L., M.C.H., E.A.N., S.H.G. and X.J.W. conceived the research. M.C.H., B.K.O., A.H.R. and M.Z. designed and executed the THz generation and characterization. E.A.N. and R.K.L. designed and fabricated the slit structure and carried out the THz-structure and particle simulations. R.K.L., M.C.H., E.A.N., B.K.O., X.S., S.P.W. and J.Y. set up the experiment and performed data collection. R.K.L., M.C.H. and E.A.N. analyzed and interpreted the data with input from A.M.L., B.K.O. and X.J.W. R.K.L., M.C.H. and E.A.N. wrote the manuscript with extensive suggestions from S.H.G. and A.M.L. and contributions from all other authors.

**Competing financial interests**

The authors declare no competing financial interests.

**Additional information**

Reprints and permissions information is available online at www.nature.com/reprints. Correspondence and requests for materials should be addressed to R.K.L. and X.J.W.

**Methods**

**Relativistic electron beams**

Experiments were conducted at the Accelerator Structure Test Area (ASTA) facility at SLAC National Accelerator Laboratory. A detailed description of the apparatus was reported in Ref [45]. For the measurement reported above, an electron beam was generated by a 2.856 GHz rf photocathode gun operated at 70 MV/m gradient. The electron bunches were generated by UV laser pulses of 100 fs FWHM pulse duration and launched at 5 degrees from the zero-crossing of the rf phase. The beam charge was 3 fC ($2 \times 10^4$ electrons) and the beam kinetic energy was 3.1 MeV. The electron beam was collimated by two focusing solenoidal lenses located at 0.19 m and 1.00 m (defined relative to the photocathode at z=0 m), respectively. The THz slit was located at 1.39 m, and the detector screen was at 4.59 m.

**THz source and characterization**

Quasi-single-cycle THz pulses were generated by optical rectification of 800 nm laser pulses in a $LiNbO_3$ crystal using the tilted pulse front method [31]. The optical pulse energy was 20 mJ and 100 fs FWHM duration. The THz pulse was focused with a 1-inch effective focal length (EFL) parabolic mirror, recollimated using a second, 3-inch diameter, 7-inch EFL parabolic mirror and then transported into the sample vacuum chamber through a polymer window. In vacuum, the THz pulse was focused by a 90 degree off-axis parabolic (OAP) mirror with 3-inch focal length toward the slit structure. A 2.5 mm diameter hole in the OAP mirror allows to collinearly overlap the relativistic electron beam, as well as an 800 nm electro-optical sampling laser pulse with the THz pulse. A 50-um thick, 110-cut GaP crystal was mounted in the same x-y plane as the slit structure. A pair of x and y translation stages were used to move the GaP crystal onto the beam axis for EO sampling characterization of the THz pulse, and to move the slit onto the beam axis for the measurement of THz streaking of the electron beams. A Spiricon Pyrocam III camera was used to the characterize the transverse spot size of the THz radiation, which was 1.5 mm at the focus. The average THz pulse energy was measured with an Ophir P3A-THz detector, however with relatively large uncertainty due to thermal fluctuations. From the magnitude of the electro-optic modulation, the THz fields strength is estimated to be 550 kV/cm. This is confirmed by matching the measured THz streaking effects with simulation.

**THz structure and field modeling**

The transverse deflecting structure consists of a rectangular slit machined out of 0.1 mm thick CuAu braze foil. The slit height is 50 microns. The slit dimension in the transverse plane perpendicular to the electric field polarization is large enough (4 mm) that the THz pulse effectively observes a parallel plate rectangular waveguide. The structure itself is therefore dispersion free, and the interaction with the electron beam is a result of the impedance mismatch between the slit and the surrounding vacuum.

To reconstruct the THz pulse interaction with the electron bunch in the transverse deflecting structure we utilize both the measured time domain EO sampling signal at the focal spot as well as the frequency domain simulations for the THz slit from HFSS. A Fourier transform of the EO sampling trace provides the spectral amplitude and phasing for the THz pulse. Frequency domain simulations are performed over a highly over-moded volume covering $\pm 4$ mm in the longitudinal direction and $\pm 2.5$ mm in the transverse direction. Simulations were performed from $0.1 - 1.5$ THz to cover the full spectrum of the input pulse. A Gaussian beam excitation was used to properly simulate the field distribution focusing on the slit. A Fourier transform is used to convert the full three-dimensional fields from the frequency domain solver with the correct amplitude and phase back into the time domain where the interaction with individual electrons is calculated.

**Evaluation of the temporal resolution for the bunch length measurement**

The spot size of the electron beams on the detector screen with the THz beam off were $52.8 \pm 1.0$ μrad rms with respect to the location of the slit. The resolution for bunch length determination, i.e. the shortest bunch length that can be evaluated, is defined as when the streaked beam size is larger than the unstreaked beam size by three times of standard deviation, i.e. larger than 55.8 μrad. At the resolution limit, the bunch length corresponds to a streaked size equal to the quadratic difference $\sqrt{55.8^2 - 52.8^2} = 18.0$ μrad. Since the streaking calibration coefficient is $D = 7.4$ μrad/fs, the resolution for bunch length determination is 2.4 fs rms. One can improve the resolution with stronger THz streaking field or minimize the unstreaked beam size by reducing the beam divergence.

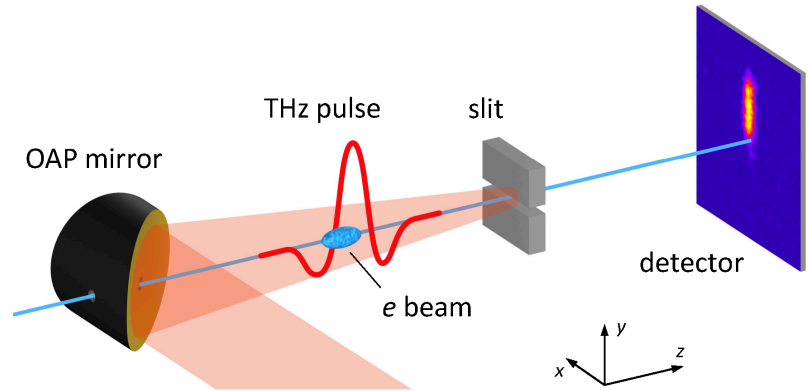

Fig. 1 Schematic of the experimental configuration. The THz radiation is focused by a 90 degree off-axis parabolic (OAP) mirror. The relativistic electron beam is injected through a hole in the OAP mirror and propagates collinearly with the THz radiation. A slit structure is located at the THz focal point and enhances the integrated momentum kick to the electron beams. The streaked electron beam profiles are recorded by a downstream imaging detector.

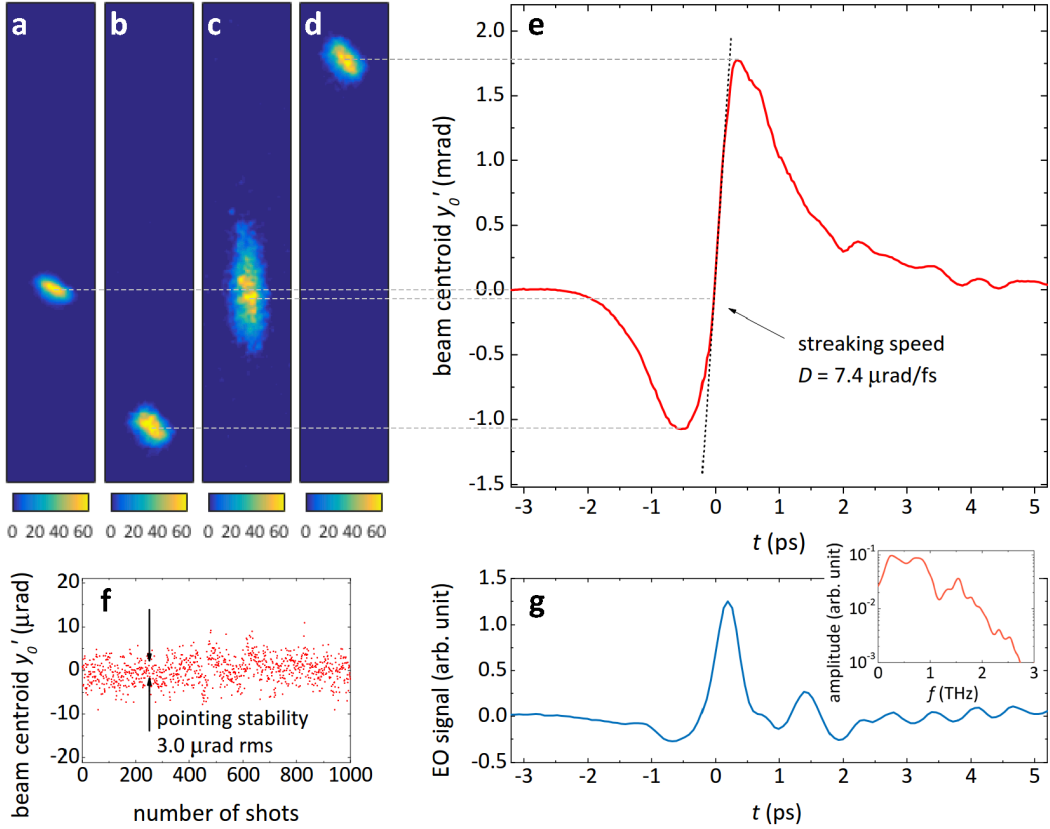

Fig. 2 Experimental data of THz streaking of relativistic electron bunches for evaluating the timing accuracy. (**a-d**) raw images of streaked electron bunches at different time delays, which are linked by dashed light gray lines to **e** showing the beam centroid in angle $y_0'$ for different time delays between the electron bunch and the THz radiation. **f** the pointing stability of the electron beam when the THz is turned off. **g** the EO sampling trace of the THz pulse. The inset shows the spectrum of the THz pulse.

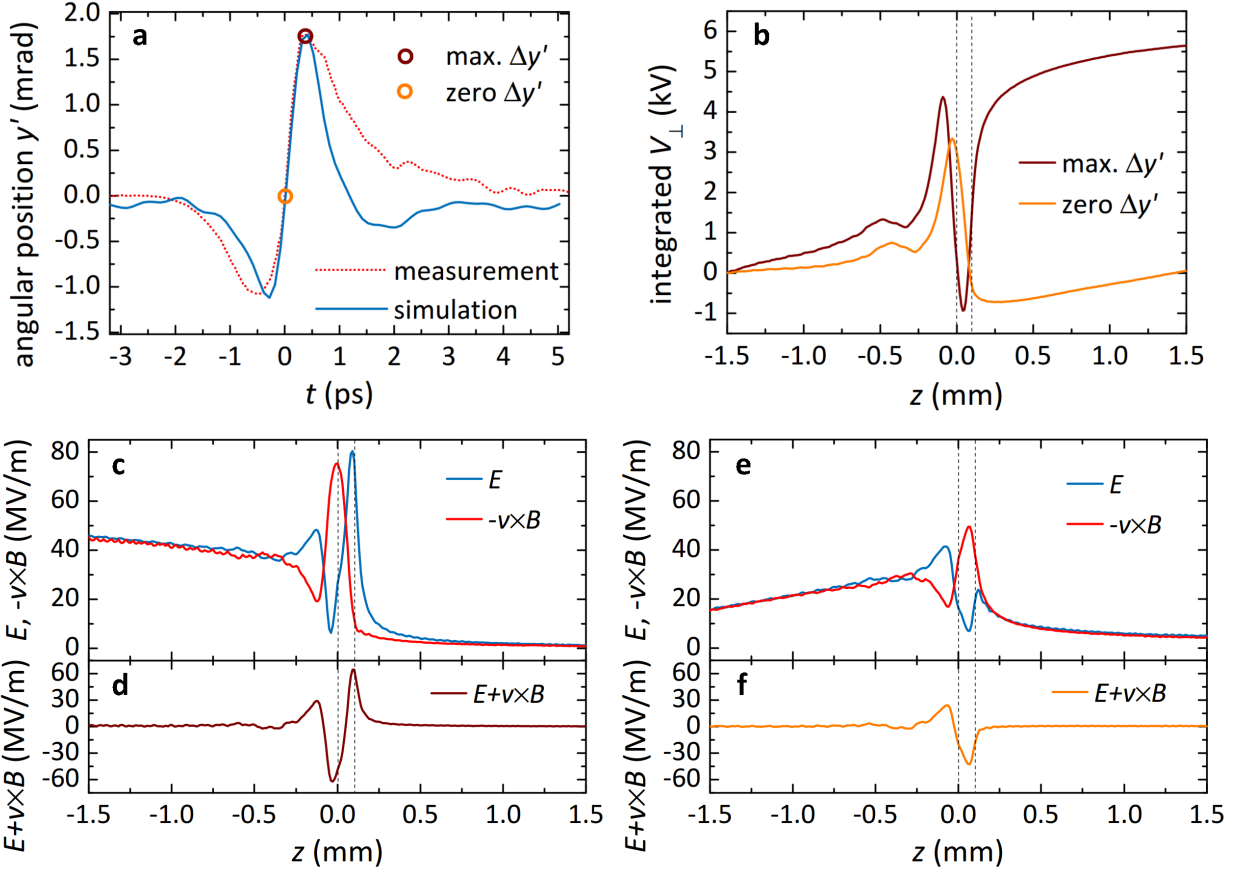

Fig. 3 Simulation results of angular position, integrated transverse voltage, and Lorentz forces experienced by the relativistic electrons under the combined influence of the THz radiation and the slit structure. **a** comparison between the measured and simulated final angular position of the electrons for different time delay. **b-f** compare the kinetics of two particles highlighted in **a**: 'max. $\Delta y'$' with the maximum change in angular position and 'zero $\Delta y'$' with the angular motion integrated to zero. In **b-f**, the horizontal axis is the coordinate of the electron as it co-propagates with the THz radiation. The focal point of the THz radiation is located at $z=0$ and the two dashed vertical lines indicate the 100-um-thick slit structure. **b** compares the integrated transverse voltage experienced by the two particles. **c** and **d** show the components and combined Lorentz force experienced by the 'max. $\Delta y'$' particle. **e** and **f** show the components and combined Lorentz force experienced by the 'zero $\Delta y'$' particle.

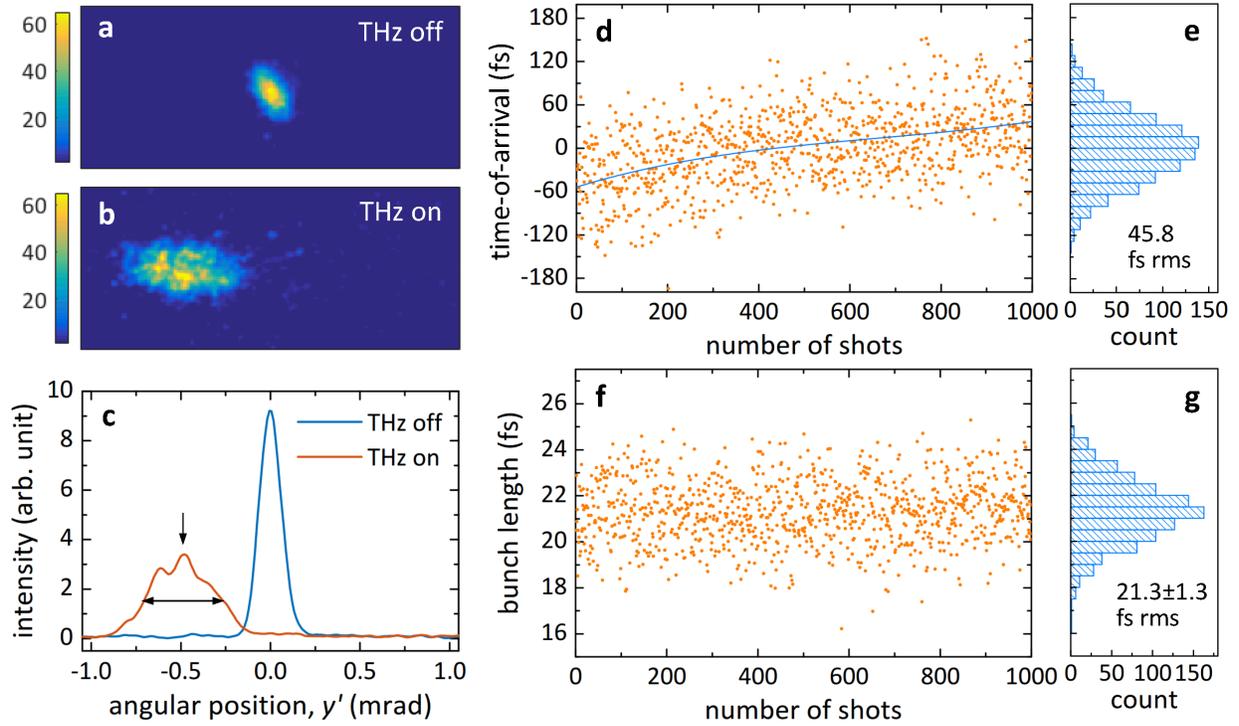

Fig. 4 Experimental data of the relative-to-laser time jitter and the bunch length of the electron beams determined from single-shot THz streaking images. From **a-c** the comparison of single-shot electron beams profiles when the THz is off and on, one can determine the relative time-of-arrival (vertical arrow in **c**) and bunch length (horizontal arrow in **c**) of every electron pulse. **d** and **e** show measurements of the timing and bunch length of 1000 consecutive electron shots. After removing the slow timing drift, the timing stability is 45.8 fs rms. **f** and **g** show the measured bunch length for the same 1000 electron shots. The bunch length is 21.3±1.3 fs rms.